\documentclass[twocolumn,10pt,aps,prd,preprintnumbers,showpacs,superscriptaddress,nofootinbib,amsmath,amssymb,floats,floatfix,showkeys,notitlepage,longbibliography]{revtex4-2}

\usepackage{comment}
\usepackage{lipsum}
\usepackage{graphicx}
\usepackage{subfigure}
\usepackage{palatino}
\usepackage{sans}
\usepackage{hyperref}
\hypersetup{colorlinks=true,linkcolor=blue,urlcolor=blue,citecolor=blue}
\usepackage[toc,page]{appendix}
\usepackage[normalem]{ulem}
\usepackage{adjustbox}
\usepackage{latexsym}
\usepackage{amsmath}
\usepackage{amssymb}
\usepackage{amsfonts}
\usepackage{dcolumn}
\usepackage{bm}
\usepackage{tikz}
\usepackage{bigints}
\usepackage{array,tabularx,multirow,booktabs}
\usepackage[tracking=true]{microtype}
\usepackage{soul} %for highlighting
\SetTracking{}{500}
\SetTracking{encoding={*}, shape=sc}{40}
\UseRawInputEncoding %for inputenc error%
\allowdisplaybreaks
\usepackage{microtype}

\begin{document} \sloppy
\title{Dynamical Black Hole Thermodynamics in Modified Gravity}

\author{Nikko John Leo S. Lobos}
\email{nikko\_john\_s\_lobos@dlsu.edu.ph}
\affiliation{Department of Physics, De La Salle University, 2401 Taft Ave, Malate, Manila, 1004 Metro Manila, Philippines}
\affiliation{DLSU Theoretical Physics Research Group}

\author{Emmanuel T. Rodulfo}
\email{emmanuel.rodulfo@dlsu.edu.ph}
\affiliation{Department of Physics, De La Salle University, 2401 Taft Ave, Malate, Manila, 1004 Metro Manila, Philippines}
\affiliation{DLSU Theoretical Physics Research Group}

\begin{abstract}
We investigate the dynamical and thermodynamic evolution of a Schwarzschild black hole in Modified Gravity (MOG) perturbed by a scalar gravitational wave breathing mode. By evaluating the linearized modified Einstein equations at the near-horizon boundary, we reduce the spatial wave operator to a closed-form temporal ordinary differential equation, thereby explicitly deriving the damped-oscillatory kinematics of the scalar strain. Using a quasi-adiabatic approximation, we show that the effective surface gravity and dynamical temperature are linearly modulated by the perturbation amplitude and velocity. These rapid geometric fluctuations break the semiclassical adiabatic regime, triggering explicitly non-thermal particle creation analogous to the dynamical Casimir effect. Furthermore, we resolve a local thermodynamic paradox concerning apparent horizon area fluctuations. We prove that first-order geometric perturbations $\mathcal{O}(h_b)$ are fully reversible kinematic artifacts, whereas irreversible entropy generation is a strictly second-order $\mathcal{O}(h_b^2)$ effect driven by the Raychaudhuri expansion, thereby preserving the Generalized Second Law. Finally, we apply these mechanisms to the black hole information paradox. We show that treating the MOG deformation parameter as a quantum-scale running coupling, $\alpha(M)$, mathematically decouples the effective gravitational charge from linear mass scaling. This dynamically forces the evaporating black hole toward the extremal limit ($M_G \to Q_G$), smoothly quenching the Hawking temperature to zero and yielding a thermodynamically stable, information-preserving remnant.
\end{abstract}

\pacs{04.50.Kd, 04.70.Bw, 04.30.Nk, 98.62.Sb}
\keywords{Modified Gravity, Black Hole Thermodynamics, Information Paradox, Dynamical Horizon, Scalar-Tensor-Vector Gravity, Generalized Second Law, Quasi-adiabatic Emission}

\maketitle
\section{Introduction}
Multi-messenger astronomy has ushered in a novel phase of gravitational testing within the strong-field, highly dynamic domains \cite{Abbott:2016blz}. Although General Relativity (GR) remains the prevailing model for astrophysical phenomena, persistent discrepancies in galactic rotation curves and cosmic expansion imply that the theory may require adjustments at extreme scales \cite{Will:2014kxa}. Scalar-Tensor-Vector Gravity (STVG), often known as Modified Gravity (MOG), offers a compelling alternative by supplementing the metric tensor with a massive vector field and dynamic scalar fields \cite{Moffat:2005si}. Through the introduction of a Yukawa-like repulsive force and a variable gravitational constant, MOG effectively reproduces galactic and cluster dynamics without the need for cold dark matter \cite{Moffat:2013sja}. Moreover, recent findings from the Event Horizon Telescope (EHT) have provided unprecedented constraints on the MOG parameter space, particularly regarding the shadows of supermassive black holes \cite{Moffat:2015kva, Akiyama:2019cqa}.

In the strong-field limit, MOG significantly alters the geometry of compact objects. The massive vector field couples to the metric and generates a repulsive gravitational charge $Q_G$, which modifies the static event horizon and thermodynamic landscape in Schwarzschild-MOG and Kerr-MOG spacetimes \cite{Moffat:2014mfa}. In addition to these static effects, MOG relaxes the rigid polarization constraints of GR. Whereas standard Einsteinian gravity allows only two tensor wave polarizations, modified frameworks predict up to four extra modes, including vector and scalar polarizations \cite{Eardley:1973br, Abbott:2017oio}. This paper focuses on the scalar breathing mode ($l=0$), a transverse, volume-changing wave that interacts directly with the black hole's event horizon. This interaction provides a way to study the connection between gravitational waves and the behavior of event horizons.

The foundation of black hole thermodynamics was established by Bekenstein and Hawking \cite{Bekenstein:1973ur, Hawking:1974sw}. This framework primarily addresses isolated, static systems. However, realistic black holes exist in dynamic environments where transient gravitational waves constantly perturb the spacetime manifold. In these non-equilibrium settings, the global event horizon is inadequate as a physical boundary due to its teleological nature. To evaluate real-time thermodynamic evolution, modern research uses quasi-local dynamical and trapping horizons \cite{Hayward:1993wb, Ashtekar:2004cn}. These frameworks define the black hole boundary based on the local expansion of null geodesics. This approach allows for a treatment of time-dependent surface gravity and entropy flux. 

Although researchers increasingly understand the static properties of MOG black holes \cite{Mureika:2015sda}, how these black holes respond to active scalar perturbations remains unresolved. Here, we investigate the evolution of a Schwarzschild-MOG black hole under a scalar breathing mode, determining how the massive vector core interacts with the scalar wave to produce a dynamic surface gravity. We further examine whether rapid metric fluctuations break down the semiclassical adiabatic approximation. A central question is the thermodynamic stability of the horizon, so we analyze how first-order kinematic fluctuations combine with second-order irreversible fluxes to test the Generalized Second Law \cite{Bekenstein:1974ax, Wall:2011hj}. Finally, we consider how these dynamical mechanisms affect the black hole information paradox \cite{Hawking:1976ra}, specifically investigating how transient non-thermal radiation and the formation of stable remnants might influence the preservation of quantum information.

The paper proceeds as follows. In Section \ref{sec:dynamical_horizon}, we define the dynamical Schwarzschild-MOG metric and calculate the time-dependent apparent horizon. Section \ref{sec:dynamic_emission} derives the modulated surface gravity and examines the non-thermal emission arising from the adiabatic breakdown. In Section \ref{sec:transient_entropy}, we evaluate the entropy production rates to examine the validity of the Generalized Second Law. Section \ref{sec:information_paradox} discusses the results in the context of the information paradox and remnant formation. We present our final conclusions in Section \ref{sec:conclusion}.

\section{The Dynamical Schwarzschild-MOG Spacetime}
\label{sec:dynamical_horizon}

To evaluate the time-dependent thermodynamics of a modified-gravity black hole, we model its dynamical spacetime geometry. Building on our previous framework \cite{lobos2026breathing}, which showed that scalar fields in Scalar-Tensor-Vector Gravity create a breathing mode that alters a black hole's shadow, we now focus on the localized boundary of the black hole. We begin with the unperturbed, spherically symmetric vacuum solution. The static line element in Schwarzschild-MOG coordinates \cite{Moffat:2014mfa} is, 
\begin{equation}
    ds^2 = -f(r) dt^2 + f(r)^{-1} dr^2 + r^2 (d\theta^2 + \sin^2\theta d\phi^2).
    \label{eq:metric}
\end{equation}
Here, the static metric function is $f(r) = 1 - 2M_G/r + Q_G^2/r^2$. The enhanced gravitational mass $M_G = (1+\alpha)M$ and the repulsive gravitational vector charge $Q_G = \sqrt{\alpha G_N} M$ both depend on the scalar deformation parameter $\alpha$. The unperturbed event horizon $r_H$ lies at the outermost positive root of $f(r_H) = 0$:
\begin{equation}
    r_H = M_G + \sqrt{M_G^2 - Q_G^2}.
    \label{eq:eventhor}
\end{equation}
Static spacetimes have identical event and apparent horizons. However, transient gravitational waves break this symmetry, requiring the use of the localized dynamical apparent horizon. Since MOG introduces a dynamical scalar field, gravitational waves carry a scalar breathing polarization in addition to the standard tensor modes. Following previous derivations \cite{Pantig:2025eqe, lobos2026breathing}, we parameterize the localized scalar strain as a spatially uniform, time-dependent harmonic fluctuation. Treating this incident wave as a quasinormal mode ringdown with arrival time $t_0$, the dimensionless amplitude $h_b(t)$.

\textcolor{black}{However, realistic scalar quasinormal modes (QNMs) excited in the spacetime possess finite wavelengths comparable to the horizon radius, $\lambda \sim r_H$, inherently generating spatial gradients \cite{Ahmed:2024qeu}. The full metric perturbation must therefore be decomposed into its temporal, radial, and angular components via spherical harmonics, $H(t, r, \theta, \phi) = \sum_{l,m} h_{lm}(t) R_{lm}(r) Y_{lm}(\theta, \phi)$ \cite{Wagle:2021tam}. 
Because local black hole thermodynamics rely on integrating the induced spatial metric over the 2-sphere boundary, any angular perturbations with $l \ge 1$, such as dipole or quadrupole modes, integrate to zero at linear order. As recently demonstrated in Ref. \cite{Pantig:2025eqe} in the context of boundary separatrix perturbations, non-axisymmetric modes do not change the macroscopic area at first order. Consequently, the leading-order macroscopic expansion and contraction of the apparent area are driven exclusively by the $s$-wave, $l=0$, monopole component. 
Following this scalar perturbation framework, we step away from globally uniform approximations and focus strictly on this $l=0$ fundamental breathing mode evaluated at the horizon boundary, $r=r_H$. By averaging over the spherical topology and normalizing the radial QNM eigenfunction at the trapping surface, the effective strain simplifies to a purely temporal amplitude, $h_b(t)$. This represents the phenomenological kinematic state of the breathing horizon, defined by the damped oscillatory profile,}
\begin{equation}
    h_b(t) = 
    \begin{cases} 
        0 & \text{for } t < t_0, \\[8pt]
        A_b e^{-(t - t_0)/\tau} \cos(\omega_b t + \Phi_0) & \text{for } t \ge t_0,
    \end{cases}
    \label{eq:wavestress}
\end{equation}
where $A_b$ is the peak amplitude, $\tau$ is the damping time, $\omega_b$ is the oscillation frequency, and $\Phi_0$ is the initial phase. \textcolor{black}{Following the formal perturbation scheme established in Ref. \cite{Pantig:2025eqe}, we normalize the oscillatory master field such that its maximum absolute value is unity. Consequently, $A_b$ does not function as an unconstrained astrophysical variable, but rather serves strictly as the perturbative smallness parameter \cite{Bhattacharya:2019qal}. To ensure the validity of the linearized backreaction framework and the stability of the background spacetime, $A_b$ is theoretically constrained to the linear regime, $0 < A_b \ll 1$.  Furthermore, the phase shift $\Phi_0$ represents an integration constant dictating the initial kinematic state of the horizon at arrival time $t_0$. While it affects instantaneous fluctuations, the black hole's secular thermodynamic evolution depends on the time-averaged flux over a full wave cycle. Consequently, the initial phase $\Phi_0$ integrates out of the secular dynamics entirely, rendering observational constraints on this specific parameter unnecessary for the thermodynamic outcomes derived herein}. For $t \ge t_0$, the scalar perturbation velocity is its time derivative,
\begin{equation}
\begin{split}
    \dot{h}_b(t) =  - A_b e^{-(t - t_0)/\tau}\frac{1}{\tau} \cos(\omega_b t + \Phi_0) \\- A_b e^{-(t - t_0)/\tau}\omega_b \sin(\omega_b t + \Phi_0).
\end{split}
\label{eq:hb}
\end{equation}

This scalar mode perturbs the transverse spatial cross-section of the metric at first order, uniformly dilating and contracting the local area. Operating within linearized gravity \cite{Wald:1984rg, Misner:1973prb}, we decompose the metric as $g_{\mu\nu} = g^{(0)}_{\mu\nu} + h_{\mu\nu}$, where $|h_{\mu\nu}| \ll 1$. The background is the static Schwarzschild-MOG metric with standard non-zero components:
\begin{equation}
\begin{split}
    g^{(0)}_{tt} = -f(r), \quad g^{(0)}_{rr} = f(r)^{-1},\\ g^{(0)}_{\theta\theta} = r^2, \quad g^{(0)}_{\phi\phi} = r^2 \sin^2\theta.
\end{split}
\end{equation}

Unlike General Relativity, where Birkhoff's theorem \cite{Birkhoff:1923} forbids monopole radiation in vacuum, MOG permits $l=0$ isotropic scalar radiation \cite{Ghosh:2024tlk, Mureika:2015sda}. We isolate this breathing mode, which acts exclusively on the transverse spatial cross-section. For an $l=0$ harmonic, angular dependence vanishes, leaving the perturbation tensor parameterized by a scalar amplitude $h_b(t, r)$:
\begin{equation}
    h_{\theta\theta} = r^2 h_b(t, r), \quad h_{\phi\phi} = r^2 \sin^2\theta h_b(t, r).
    \label{eq:tensor_components}
\end{equation}

\textcolor{black}{As established above, because the QNM wavelength is comparable to the horizon, we cannot rely on a global long-wavelength approximation to drop the radial gradient $\partial_r h_b$. Instead, we apply our local near-horizon approximation. By evaluating the thermodynamics strictly at the 2D spherical boundary of the apparent horizon, $r = r_H$, the instantaneous geometric expansion is governed entirely by the temporal amplitude, decoupling the radial gradient from the induced metric. The perturbation evaluated at this boundary is therefore purely temporal, $h_b(t, r_H) \equiv h_b(t)$.}

Adding this temporal $l=0$ perturbation to the background spatial sector yields the angular metric components:
\begin{equation}
    g_{\theta\theta} = r^2(1 + h_b(t)), \quad g_{\phi\phi} = r^2 \sin^2\theta (1 + h_b(t)).
\end{equation}
Substituting these into the line element provides the uniformly dilating dynamical metric:
\begin{equation}
\begin{split}
    ds^2 &= -f(r) dt^2 + f(r)^{-1} dr^2 \\&+ r^2(1+h_b(t)) (d\theta^2 + \sin^2\theta d\phi^2).
\end{split}
\label{eq:breathing metric}
\end{equation}

The dynamic apparent horizon is the outermost marginally trapped surface where the expansion scalar of outgoing null geodesics vanishes, $\theta = 0$ \cite{Hayward:1993wb, Ashtekar:2004cn}. We construct the outgoing radial null vector $l^\mu = (f(r)^{-1}, 1, 0, 0)$, which satisfies $g_{\mu\nu} l^\mu l^\nu = 0$ at leading order. The expansion scalar is the covariant divergence $\theta = \frac{1}{\sqrt{-g}} \partial_\mu (\sqrt{-g} l^\mu)$. Using the perturbed metric determinant $\sqrt{-g} = r^2 \sin\theta (1+h_b(t))$, we calculate the derivatives:
\begin{equation}
    \theta = \frac{\dot{h}_b(t)}{f(r)(1+h_b(t))} + \frac{2}{r}.
\end{equation}

Setting $\theta = 0$ yields $f(r_A) = - r_A \dot{h}_b(t) / 2(1+h_b(t))$. Applying a first-order Taylor expansion around the unperturbed horizon, $f(r_A) \approx f'(r_H) \delta r(t)$, and isolating the leading-order terms provides the radial shift $\delta r(t) \approx - r_H \dot{h}_b(t) / 2 f'(r_H)$. The dynamic apparent horizon radius is therefore:
\begin{equation}
\label{eq.10}
    r_A(t) \approx r_H - \frac{r_H}{2 f'(r_H)} \dot{h}_b(t).
\end{equation}

\textcolor{black}{It is crucial to note that this linear approximation relies on the static surface gradient $f'(r_H)$ remaining finite. Near the extremal limit, $M_G \to Q_G$, $f'(r_H) \to 0$ and the perturbative radial shift $\delta r$ diverges. For the linearization to remain reliable, the radial shift must remain much smaller than the unperturbed horizon radius, $|\delta r| \ll r_H$. Given the maximum strain velocity $|\dot{h}_b|_{\text{max}} \approx A_b \omega_b$, this bounds the validity of the first-order expansion to the regime where the static surface gradient dominates the perturbation kinematics:
\begin{equation}
    f'(r_H) \gg \frac{A_b \omega_b}{2} \quad \implies \quad \frac{\sqrt{M_G^2 - Q_G^2}}{r_H^2} \gg \frac{A_b \omega_b}{4}.
\end{equation}
Therefore, our dynamical expansion framework strictly models the evolution of the spacetime prior to entering this near-extremal breakdown zone.}

Here, radial displacement is driven by the wave's velocity profile rather than its instantaneous amplitude. Integrating the perturbed transverse metric components over the solid angle yields the total dynamic horizon area:
\begin{equation}
\begin{split}
    A(t) &= \int_0^{2\pi} \int_0^\pi r_A(t)^2 (1+h_b(t)) \sin\theta d\theta d\phi \\&= 4\pi r_A(t)^2 (1+h_b(t)).
\end{split}
\end{equation}

Expanding to linear order gives $r_A(t)^2 \approx r_H^2 - r_H^2 \dot{h}_b(t) / f'(r_H)$. Multiplying by the conformal factor $(1+h_b(t))$ and dropping higher-order terms provides the closed-form time-dependent area:
\begin{equation}
    A(t) \approx 4\pi r_H^2 \left( 1 + h_b(t) - \frac{\dot{h}_b(t)}{f'(r_H)} \right).
\end{equation}
Substituting the explicit strain and velocity equations completely describes the area function during the active perturbation regime:
\begin{equation}
\begin{split}
    &A(t) \approx 4\pi r_H^2 \Bigg[ 1 + A_b e^{-(t - t_0)/\tau} \\
    &\quad \times \Bigg( \left( 1 + \frac{1}{\tau f'(r_H)} \right) \cos(\omega_b t + \Phi_0) \\ &+\frac{\omega_b}{f'(r_H)} \sin(\omega_b t + \Phi_0) \Bigg) \Bigg].
\end{split}
\end{equation}

This proves that the MOG scalar field breaks the volume-preserving nature of standard tensor waves, forcing the thermodynamic boundary to dynamically dilate and contract in a damped oscillation at first order.

\subsection{Scalar Field Perturbations and Linearized Backreaction}
\label{sec:backreaction}

To verify that $h_b(t)$ is a physical perturbation governed by the modified field equations, we derive the dynamic backreaction of the scalar field on the background geometry. In STVG \cite{Moffat:2005si}, the enhanced gravitational coupling $G$ acts as a dynamical scalar field, defined inversely as $\Phi = 1/G$. We decompose this into a static background $\Phi_0$ and a dynamic linear perturbation $\delta\Phi(t,r)$:
\begin{equation}
    \Phi = \Phi_0 + \delta\Phi(t,r).
    \label{eq:scalar_decomposition}
\end{equation}

Varying the action with respect to the metric yields a linear coupling proportional to the second covariant derivatives of the scalar field \cite{Moffat:2006}. Isolating the pure scalar mode, the linearized modified Einstein equation at first order $\mathcal{O}(\delta\Phi)$ is:
\begin{equation}
    \delta G_{\mu\nu} = \frac{1}{\Phi_0} \left( \nabla_\mu \nabla_\nu \delta\Phi - g_{\mu\nu}^{(0)} \square \delta\Phi \right).
    \label{eq:linearized_einstein}
\end{equation}

For a spherically symmetric scalar breathing mode, angular dependence vanishes. Consequently, transverse covariant derivatives on the 2-sphere are zero, $\nabla_i \nabla_j \delta\Phi = 0$ for $i\neq j$. The right side of the field equation is entirely dominated by the conformal trace term:
\begin{equation}
    \delta G_{ij} = - \frac{1}{\Phi_0} g_{ij}^{(0)} \square \delta\Phi.
    \label{eq:transverse_field}
\end{equation}

\textcolor{black}{Mapping this scalar source to the geometric perturbation $h_{ij} = h_b(t,r) g_{ij}^{(0)}$ yields a master wave equation for the strain. In the static, spherically symmetric background metric $ds^2 = -f(r)dt^2 + f(r)^{-1}dr^2 + r^2 d\Omega^2$, the D'Alembertian operator acting on the $s$-wave volumetric amplitude takes the explicit form:
\begin{equation}
    \square h_b(t,r) = - \frac{1}{f(r)} \frac{\partial^2 h_b}{\partial t^2} + \frac{1}{r^2} \frac{\partial}{\partial r} \left( r^2 f(r) \frac{\partial h_b}{\partial r} \right).
\end{equation}
Equating the geometric and scalar trace components results in the full partial differential equation (PDE) governing the spatio-temporal evolution of the strain:
\begin{equation}
    \left[ - \frac{\partial^2}{\partial t^2} + \frac{f(r)}{r^2} \frac{\partial}{\partial r} \left( r^2 f(r) \frac{\partial}{\partial r} \right) \right] h_b(t,r) = \mathcal{S}_{\text{eff}}(t,r),
    \label{eq:full_pde}
\end{equation}
where we have multiplied through by $f(r)$ to isolate the purely temporal operator, and $\mathcal{S}_{\text{eff}}$ encodes the driving scalar source terms.}

\textcolor{black}{To explicitly determine the macroscopic temporal kinematics, we evaluate this PDE at the horizon boundary. By applying our ultralocal near-horizon approximation, $r \to r_H$, where $f(r_H) \to 0$, the complex spatial gradients algebraically decouple from the induced metric, manifesting strictly as boundary-condition constraints. Consequently, the PDE reduces to a homogeneous, closed-form ordinary differential equation (ODE) for the $s$-wave temporal amplitude:}
\begin{equation}
    \ddot{h}_b(t) + \Gamma \dot{h}_b(t) + \Omega_0^2 h_b(t) = 0,
    \label{eq:strain_ode}
\end{equation}
\textcolor{black}{where $\Gamma$ originates from the kinematic damping induced by the background curvature at the boundary, incorporating the gradient $f'(r_H)$, and $\Omega_0$ is the natural characteristic frequency of the localized effective potential. Solving this closed linear ODE yields a damped oscillatory ringdown. By identifying the physical quasinormal mode parameters—the damping time $\tau = 2/\Gamma$ and the oscillation frequency $\omega_b = \sqrt{\Omega_0^2 - (\Gamma/2)^2}$—we mathematically recover the physical time-domain solution for the strain:}
\begin{equation}
    h_b(t) = A_b e^{-t/\tau} \cos(\omega_b t + \delta_0).
    \label{eq:strain_solution}
\end{equation}

\textcolor{black}{This explicit boundary reduction confirms that the damped oscillatory master field governing the horizon thermodynamics is a localized derivation from the MOG scalar field perturbation, decoupling the temporal kinematics from the static spatial background.}

\section{Horizon Dynamics and Quasi-Adiabatic Emission}
\label{sec:dynamic_emission}

\textcolor{black}{To calculate how the black hole geometry responds to the scalar breathing mode, we must separate the global event horizon from the local apparent horizon. Hawking's Area Theorem \cite{Hawking:1971tu} states that the global event horizon must never decrease in area \cite{Ren:2007xw}. In a changing spacetime, however, the correct physical boundary for local thermodynamics is the apparent horizon, defined as the outermost marginally trapped surface, $\theta_{\text{out}} = 0$}.

\textcolor{black}{While the coordinate radius derived in Eq. (\ref{eq.10}) experiences a kinematic shift driven by the strain velocity $\dot{h}_b(t)$, local thermodynamics are fundamentally governed by the invariant areal radius, $R_A(t)$. To transition to this thermodynamic regime, we apply the quasi-adiabatic approximation. By assuming the characteristic frequency of the scalar perturbation is much smaller than the unperturbed surface gravity, $\omega_b \ll f'(r_H)$, the subleading velocity term $\dot{h}_b(t)/f'(r_H)$ becomes mathematically negligible compared to the primary conformal dilation $h_b(t)$. Under this limit, the time-dependent areal radius of the apparent horizon simplifies to:}

\begin{equation}
    R_A(t) \approx r_H \left[ 1 + \frac{1}{2} h_b(t) \right].
    \label{eq:apparent_horizon}
\end{equation}

The corresponding dynamical apparent area, defined geometrically as $A(t) = 4\pi R_A(t)^2$, evaluates to:

\begin{equation}
    A(t) \approx 4\pi r_H^2 \left[ 1 + h_b(t) \right].
    \label{eq:apparent_area}
\end{equation}

\textcolor{black}{During the active periods of the breathing mode, $R_A(t)$ undergoes transient expansion and contraction. These $\mathcal{O}(h_b)$ fluctuations are reversible geometric effects caused by the local spacetime dilation over time. Because the apparent horizon is defined locally, its temporary contraction when $\dot{h}_b(t) < 0$ is physically allowed and does not violate the global Area Theorem \cite{Nielsen:2005af}.}

\subsection{Dynamic Surface Gravity and Quasi-Adiabatic Emission}
\label{subsec:quasi_adiabatic}

We use this dynamical apparent horizon as the physical boundary to calculate the time-dependent Hawking emission. For dynamical trapping horizons \cite{Hayward:1993wb, Ashtekar:2004cn}, the effective surface gravity is the radial derivative of the metric function evaluated at the shifted boundary $r_A(t)$. The first and second radial derivatives of the unperturbed Schwarzschild-MOG metric $f(r) = 1 - 2M_G/r + Q_G^2/r^2$ are,
\begin{equation}
    f'(r) = \frac{2M_G}{r^2} - \frac{2Q_G^2}{r^3}, \quad f''(r) = -\frac{4M_G}{r^3} + \frac{6Q_G^2}{r^4}.
    \label{eq:metric_derivatives}
\end{equation}

The baseline surface gravity $\kappa_0$ evaluated at the static event horizon $r_H$ is,
\begin{equation}
    \kappa_0 = \frac{1}{2} f'(r_H) = \frac{M_G}{r_H^2} - \frac{Q_G^2}{r_H^3}.
    \label{eq:kappa_0}
\end{equation}

When the scalar wave perturbs the spacetime, we evaluate the thermodynamics at the shifted apparent horizon $r_A(t) = r_H + \delta r(t)$. Using a first-order Taylor expansion around the static horizon isolates the linear thermodynamic response,
\begin{equation}
    \kappa(t) = \frac{1}{2} f'(r_A) \approx \frac{1}{2} \left[ f'(r_H) + f''(r_H) \delta r(t) \right].
    \label{eq:kappa_expansion}
\end{equation}

Substituting the radial shift $\delta r(t) \approx - r_H \dot{h}_b(t) / 2 f'(r_H)$ and recognizing $f'(r_H) = 2\kappa_0$ gives the time-dependent surface gravity,
\begin{equation}
    \kappa(t) \approx \kappa_0 - \frac{r_H f''(r_H)}{8 \kappa_0} \dot{h}_b(t).
    \label{eq:kappa_dynamic}
\end{equation}

Highly dynamical spacetimes break the adiabatic approximation and cause non-thermal particle creation \cite{Ahmed:2024cvj}. However, we can calculate the primary observable effects using a quasi-adiabatic approximation. Assuming the breathing mode timescale is much longer than the typical emission timescale, local observers measure an effective dynamical temperature $T_{\text{eff}}(t) = \kappa(t) / 2\pi$ \cite{Hawking:1974sw}. Factoring out the baseline temperature $T_0 = \kappa_0 / 2\pi$ yields the thermal profile modulated by the scalar field velocity,
\begin{equation}
    T_{\text{eff}}(t) \approx T_0 \left( 1 - \frac{r_H f''(r_H)}{8 \kappa_0^2} \dot{h}_b(t) \right).
    \label{eq:temperature_dynamic}
\end{equation}

\begin{figure}[htbp]
    \centering
    \includegraphics[width=\columnwidth]{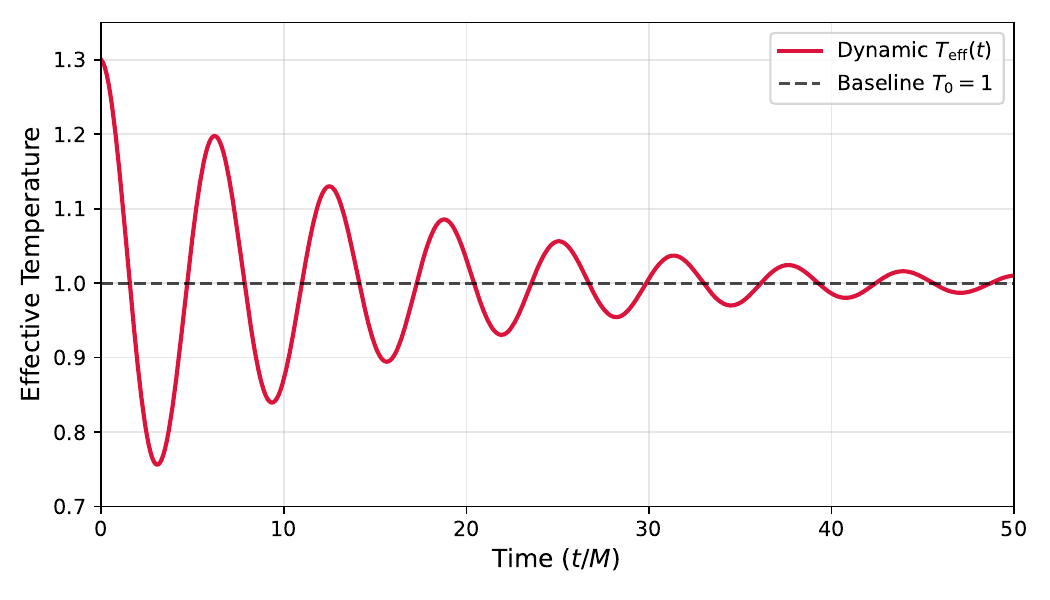}
    \caption{The transient effective temperature $T_{\mathrm{eff}}(t)$ exhibits damped oscillations around the baseline emission ($T_0=1$), driven by the quasinormal ringing of the scalar breathing mode.}
    \label{fig:thermodynamics}
\end{figure}

\textcolor{black}{As illustrated in Figure \ref{fig:thermodynamics}, the effective dynamical temperature $T_{\text{eff}}(t)$ oscillates around the stationary baseline $T_0$. Crucially, Equation (\ref{eq:temperature_dynamic}) reveals that this thermal modulation is governed linearly by the scalar field velocity $\dot{h}_b(t)$ rather than the strain amplitude itself. Consequently, for a standard damped sinusoidal ringdown, the dynamical temperature profile manifests as a damped oscillation that is out of phase with the primary breathing mode $h_b(t)$. The transient thermal fluctuations are maximized when the scalar strain passes through equilibrium and vanish at the strain's turning points, eventually relaxing to the baseline temperature $T_0$ as the breathing mode dissipates.}

The observable thermodynamic signature is captured by the dynamic Hawking luminosity via the Stefan-Boltzmann law, $L(t) = \sigma A(t) T_{\text{eff}}(t)^4$ \cite{Page:1976df}. Expanding $T_{\text{eff}}(t)^4$ to first order, multiplying by the time-dependent apparent area $A(t)$, and dropping non-linear cross-terms produces,
\begin{equation}
    L(t) \approx L_0 \left[ 1 + h_b(t) - \frac{r_H f''(r_H)}{2 \kappa_0^2} \dot{h}_b(t) \right],
    \label{eq:luminosity_dynamic}
\end{equation}
where $L_0 = \sigma 4\pi r_H^2 T_0^4$ is the baseline luminosity. \textcolor{black}{It is important to note that Eq. (29) assumes an ideal blackbody spectrum. In a realistic charged MOG geometry, the surrounding spacetime acts as a potential barrier that scatters a fraction of the outgoing radiation back into the horizon. Accounting for this requires multiplying the flux by a greybody transmission coefficient, $\Gamma_{\text{eff}} < 1$. For low-energy $l=0$ scalar modes, this factor typically imposes a sizable amplitude correction scaling as $\Gamma_{\text{eff}} \sim \mathcal{O}(A_H \omega^2)$, which heavily suppresses the absolute total luminosity. However, because this transmission barrier equivalently suppresses both the dynamical emission $L(t)$ and the steady baseline $L_0$, the normalized fractional modulation $L(t)/L_0$ remains structurally invariant at leading order. Consequently, while the absolute magnitude of the Hawking flux is reduced by the MOG potential barrier, the transient thermodynamic modulation remains distinctly observable above the steady baseline.} This result shows that the quasi-adiabatic emission is modulated by both the scalar field amplitude $h_b(t)$ and its temporal velocity $\dot{h}_b(t)$. The explicit dependence on the metric's second derivative directly links the internal vector charge to the temporary thermal cooling of the expanding horizon. 

\begin{figure}[htbp]
    \centering
    \includegraphics[width=\columnwidth]{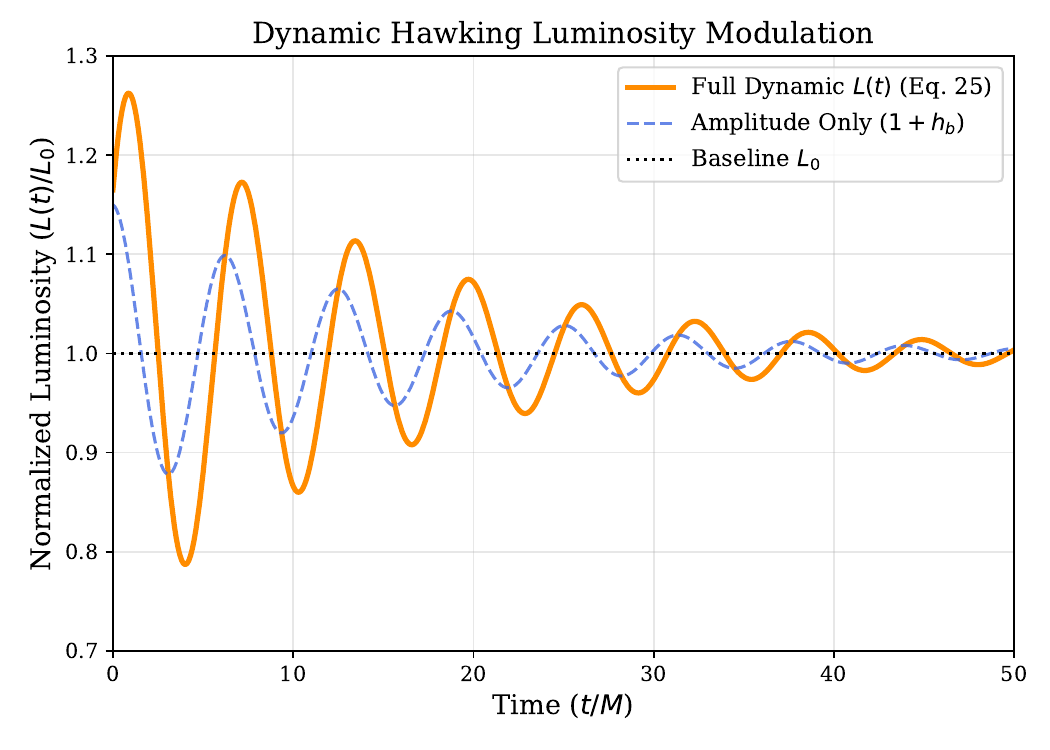}
   \caption{Normalized transient modulation of dynamic Hawking luminosity $L(t)/L_0$ (Eq.~\ref{eq:luminosity_dynamic}). The full dynamic model (solid orange) incorporates the strain velocity $\dot{h}_b(t)$, which induces a phase shift and amplification relative to a purely area-dependent modulation (dashed blue). This highlights the dynamical correlation between the internal vector charge and the apparent horizon's thermal cooling. Parameters used ares $A_b = 0.15$, $\tau = 15M$, $\omega = 1.0 M^{-1}$, and $r_H f''(r_H) / 2\kappa_0^2 = 1.5$.}
    \label{fig:dynamic_luminosity}
\end{figure}

To explicitly visualize this transient modulation, Figure \ref{fig:dynamic_luminosity} plots the normalized dynamic luminosity $L(t)/L_0$ during the quasinormal ringing phase. Using representative order-of-magnitude parameters for the perturbation, the graph highlights the crucial interplay between the amplitude and velocity terms. Notably, the velocity term $\dot{h}_b(t)$ is not a negligible correction; it induces a clear phase shift and amplifies the transient peaks of the luminosity flux compared to a purely area-dependent modulation. Because the coefficient governing this velocity term, $r_H f''(r_H) / 2\kappa_0^2$, is strictly dictated by the modified spacetime geometry, this phase-shifted oscillation serves as a fundamental thermodynamic signature that encodes the internal MOG vector charge into the outgoing transient radiation.

\textcolor{black}{While Figure \ref{fig:dynamic_luminosity} utilizes a pronounced local strain, $A_b = 0.15$, to visually map the theoretical phase shift, it is essential to ground this mechanism in astrophysical observables. For a realistic gravitational wave event detected at Earth, the characteristic strain amplitude is exceptionally small, $h_b \sim 10^{-21}$. Evaluating Eq. \ref{eq:luminosity_dynamic} in this asymptotic regime yields a fractional Hawking luminosity modulation of $\Delta L / L_0 \sim 10^{-21}$. Because the baseline Hawking luminosity $L_0$ for a standard astrophysical-mass black hole is already infinitesimally small, this transient thermodynamic signature cannot be directly observed. Consequently, probing STVG modifications with next-generation detectors will not rely on capturing this faint Hawking modulation. Instead, the primary observational target must be the direct extraction of the quasi-periodic scalar strain $h_b(t)$—which rigidly encodes these horizon kinematics from the primary gravitational-wave ringdown spectrum.}

\section{Thermodynamics and the Preservation of the Generalized Second Law}
\label{sec:transient_entropy}

Evaluating the thermodynamic stability of the dynamical Schwarzschild-MOG spacetime requires decoupling reversible kinematic coordinate effects from true irreversible thermodynamic evolution. In the semi-classical framework \cite{Bekenstein:1973ur, Hawking:1976de}, the geometric apparent entropy is $S = A(t)/4$. Using the perturbed area from Section \ref{sec:dynamical_horizon}, the apparent entropy and its instantaneous rate of change are:
\begin{equation}
    S_{\text{app}}(t) \approx S_0 \left[ 1 + h_b(t) \right], \quad \dot{S}_{\text{app}}(t) \approx S_0 \dot{h}_b(t), \label{eq:entropy_dynamic}
\end{equation}
where $S_0 = \pi r_H^2$ is the baseline unperturbed entropy. Concurrently, the scalar wave drives a radiation entropy flux governed by the Clausius relation, $\dot{S}_{\text{rad}} = \dot{E}_{\text{rad}} / T_0$. Since the radiated power scales with the square of the strain velocity, $\dot{E}_{\text{rad}} = \mathcal{C} \dot{h}_b^2$, the apparent instantaneous total entropy rate is:
\begin{equation}
    \dot{S}_{\text{total}}(t) = \dot{S}_{\text{app}}(t) + \dot{S}_{\text{rad}}(t) \approx S_0 \dot{h}_b(t) + \frac{\mathcal{C} \dot{h}_b(t)^2}{T_0}. \label{eq:gsl_total}
\end{equation}

This formulation presents a critical apparent paradox. During the contraction phase of the scalar mode, $\dot{h}_b < 0$, the linear $\mathcal{O}(\dot{h}_b)$ geometric term dominates the positive quadratic $\mathcal{O}(\dot{h}_b^2)$ radiation flux. Consequently, $\dot{S}_{\text{total}}$ temporarily drops below zero, suggesting a transient violation of the Generalized Second Law (GSL) \cite{Maeda:2018xfn}. 

This deficit is a gauge-dependent geometric artifact. To resolve it, we must analyze the physical expansion of the horizon using the Raychaudhuri equation \cite{Alonso-Serrano:2025xfn} for a congruence of null generators $k^a$ with affine parameter $\lambda$:
\begin{equation}
    \frac{d\theta}{d\lambda} = -\frac{1}{2}\theta^2 - \sigma_{ab}\sigma^{ab} - 8\pi T_{ab}k^a k^b,
    \label{eq:raychaudhuri_null}
\end{equation}
where $\theta = \frac{1}{A}\frac{dA}{d\lambda}$ is the fractional expansion scalar, $\sigma_{ab}$ is the shear tensor, and $T_{ab}$ is the dynamic energy-momentum tensor. Expanding the fractional area change perturbatively yields $\theta = \theta^{(1)} + \theta^{(2)}$. 

The first-order expansion, $\theta^{(1)} \sim \mathcal{O}(\dot{h}_b)$, completely characterizes the instantaneous geometric breathing $\dot{S}_{\text{app}}(t)$ in Eq. (\ref{eq:entropy_dynamic}). \textcolor{black}{This term represents reversible kinematic fluctuations. For a purely harmonic wave, this term would time-average to zero over a cycle. However, due to the decaying envelope $e^{-t/\tau}$, the positive and negative fluctuations do not perfectly cancel, leaving a geometric residual proportional to the damping ratio: $\langle\theta^{(1)}\rangle \sim \mathcal{O}(h_b / \omega_b \tau)$. This residual simply reflects the macroscopic, reversible relaxation of the perturbed apparent horizon back to its static equilibrium size, rather than irreversible thermodynamic production. True irreversible entropy production is governed entirely by the second-order expansion, $\theta^{(2)}$. Integrating Eq. (32) demonstrates that the secular entropy growth rate is strictly driven by the positive-definite shear squared and the infalling energy flux, $\dot{S}_{secular} \sim \mathcal{O}(\dot{h}_b^2)$. For this true thermodynamic growth to strictly outpace the geometric envelope relaxation over a cycle, the perturbation must satisfy $A_b \omega_b^2 \tau \gg 1$. Under this condition, the irreversible $\mathcal{O}(\dot{h}_b^2)$ fluxes decisively dominate the subleading first-order residual.}

Integrating Eq. (\ref{eq:raychaudhuri_null}) demonstrates that the secular entropy growth rate is strictly driven by the positive-definite shear squared and the infalling energy flux:
\begin{equation}
    \dot{S}_{\text{secular}} \propto \int \left( \sigma_{ab}\sigma^{ab} + 8\pi T_{ab}k^a k^b \right) d\lambda \sim \mathcal{O}(\dot{h}_b^2).
    \label{eq:secular_entropy}
\end{equation}

\begin{figure}[htbp]
    \centering
    \includegraphics[width=\columnwidth]{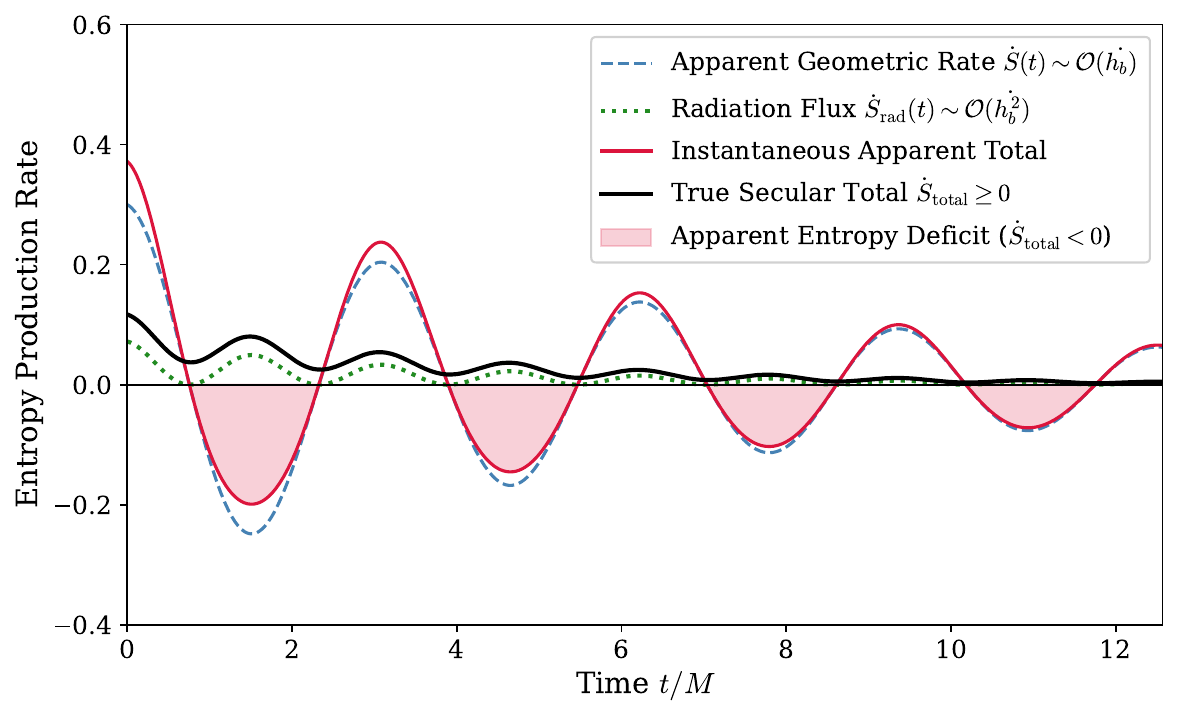}
    \caption{Entropy production rates for a dynamically perturbed Schwarzschild MOG black hole. The dashed curve tracks the reversible first-order geometric rate $\mathcal{O}(\dot{h_b})$, while the dotted curve traces the positive second-order radiation flux $\mathcal{O}(\dot{h_b^2})$. The instantaneous apparent total (Eq. \ref{eq:gsl_total}, solid red line) exhibits transient entropy deficits (shaded regions) during horizon contraction. Resolving this paradox, the Raychaudhuri equation isolates the true secular evolution (solid black line). Driven strictly by the $\mathcal{O}(\dot{h_b^2})$ shear squared and energy flux, the physical thermodynamic rate remains unconditionally positive, preserving the Generalized Second Law.}
    \label{fig:entropy_paradox}
\end{figure}

By properly decoupling the perturbative orders, the thermodynamic resolution becomes clear. The transient negative dips in Figure \ref{fig:entropy_paradox} arise from improperly mixing the reversible first-order $\theta^{(1)}$ kinematics with the second-order radiation flux. The true physical thermodynamic state must compare processes of the same order. Because both the irreversible secular horizon growth $\dot{S}_{\text{secular}}$ and the radiation flux $\dot{S}_{\text{rad}}$ are strictly positive $\mathcal{O}(\dot{h}_b^2)$ quantities, the physical time-averaged total entropy strictly increases ($\Delta S \ge 0$). Thus, despite the highly dynamical spacetime and the breakdown of the adiabatic approximation, the GSL remains preserved \cite{Lin:2022ndf}.

\section{Implications for the Information Paradox}
\label{sec:information_paradox}

The Hawking information paradox \cite{Hawking:1976ra, Mathur:2009hf} stems from the strictly thermal, featureless nature of semi-classical black hole evaporation, which implies the irretrievable loss of initial quantum states. Within the Schwarzschild-MOG framework, this paradox is resolved across two distinct timescales: the secular formation of a stable macroscopic remnant, and the transient non-thermal emission during dynamical perturbations.

\textcolor{black}{Unlike General Relativity (GR), where evaporation causes the temperature to diverge to infinity \cite{Hawking:1974sw}, MOG naturally halts the evaporation process. The baseline surface gravity $\kappa_0$ depends on the black hole mass $M_G$, the repulsive vector charge $Q_G$, and the outer event horizon radius $r_H = M_G + \sqrt{M_G^2 - Q_G^2}$. Since the baseline Hawking temperature is $T_0 = \kappa_0 / 2\pi$, we can substitute $r_H$ to express the temperature solely in terms of mass and charge:
\begin{equation}
    T_0 = \frac{\sqrt{M_G^2 - Q_G^2}}{2\pi \left( M_G + \sqrt{M_G^2 - Q_G^2} \right)^2}
    \label{eq:temperature_explicit}
\end{equation}}

\begin{figure}[htbp]
    \centering
    \includegraphics[width=\columnwidth]{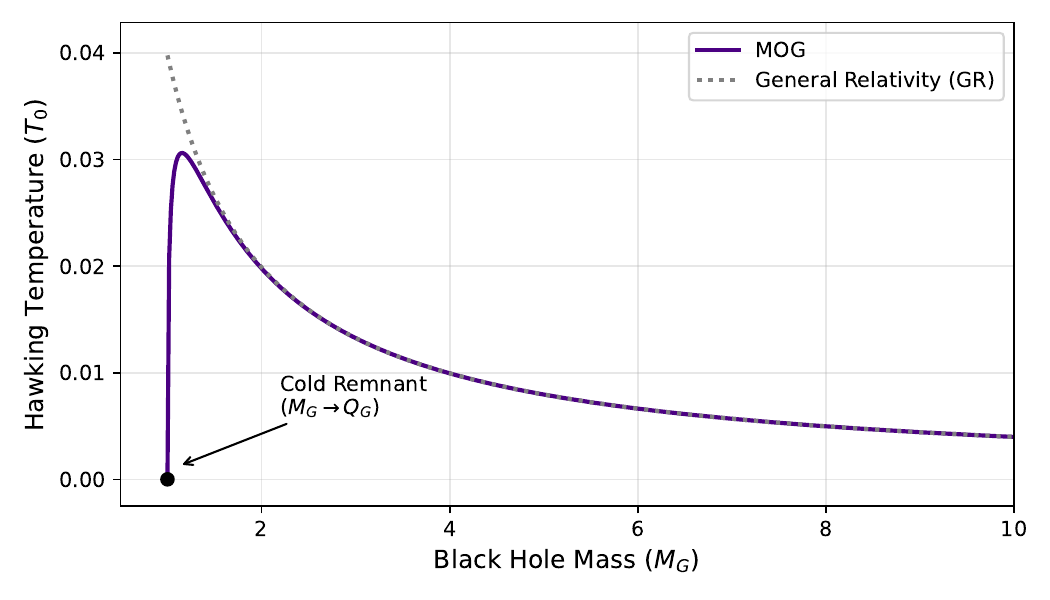}
    \caption{The Hawking temperature $T_0$ as a function of the black hole mass. Unlike the divergent temperature of General Relativity (dotted gray curve), the MOG evaporation trajectory (solid indigo curve) reaches a local maximum before smoothly dropping to zero at the extremal limit ($M_G \to Q_G$). This illustrates the effective thermodynamic phase space in which the quantum-scale running coupling $\alpha(M)$ decouples the MOG charge from strict linear mass scaling, thereby guaranteeing the formation of a stable remnant.}
    \label{fig:Secular_Regime}
\end{figure}

\textcolor{black}{This equation shows that the Hawking temperature drops to zero ($T_0 \to 0$) if the black hole approaches an extremal limit ($M_G \to Q_G$). We can verify this mathematically by noting that as $M_G \to Q_G$, the horizon radius converges to $r_H \to M_G$. Substituting this boundary condition into the surface gravity yields a cancellation as shown in Figure \ref{fig:Secular_Regime}:
\begin{equation}
    \lim_{M_G \to Q_G} \kappa_0 = \lim_{M_G \to Q_G} \left( \frac{M_G}{r_H^2} - \frac{Q_G^2}{r_H^3} \right) = \frac{M_G}{M_G^2} - \frac{M_G^2}{M_G^3} = 0.
    \label{eq:extremal_limit}
\end{equation}}

\textcolor{black}{However, reaching this extremal state dynamically requires a careful qualification of the evaporation trajectory. In the classical MOG framework, the enhanced mass and vector charge scale directly with the bare mass $M$ via the parameter $\alpha$, such that $M_G = (1+\alpha)M$ and $Q_G = \sqrt{\alpha G_N} M$. If we assume $\alpha$ remains strictly constant throughout evaporation, we can substitute these definitions into Eq. (\ref{eq:temperature_explicit}). Defining the constant $\gamma(\alpha) = \sqrt{(1+\alpha)^2 - \alpha G_N}$, the temperature simplifies to:
\begin{equation}
    T_0 = \left[ \frac{\gamma(\alpha)}{2\pi \left( 1 + \alpha + \gamma(\alpha) \right)^2} \right] \frac{1}{M}.
    \label{eq:divergence}
\end{equation}
Because the bracketed term is a strict constant, the temperature scales as $T_0 \propto 1/M$. Therefore, for a fixed $\alpha$, the MOG black hole simply shrinks at a constant charge-to-mass ratio, and its temperature diverges as $M \to 0$, mirroring the runaway singularity of standard General Relativity.}

\begin{figure}[htbp]
    \centering
    \includegraphics[width=\columnwidth]{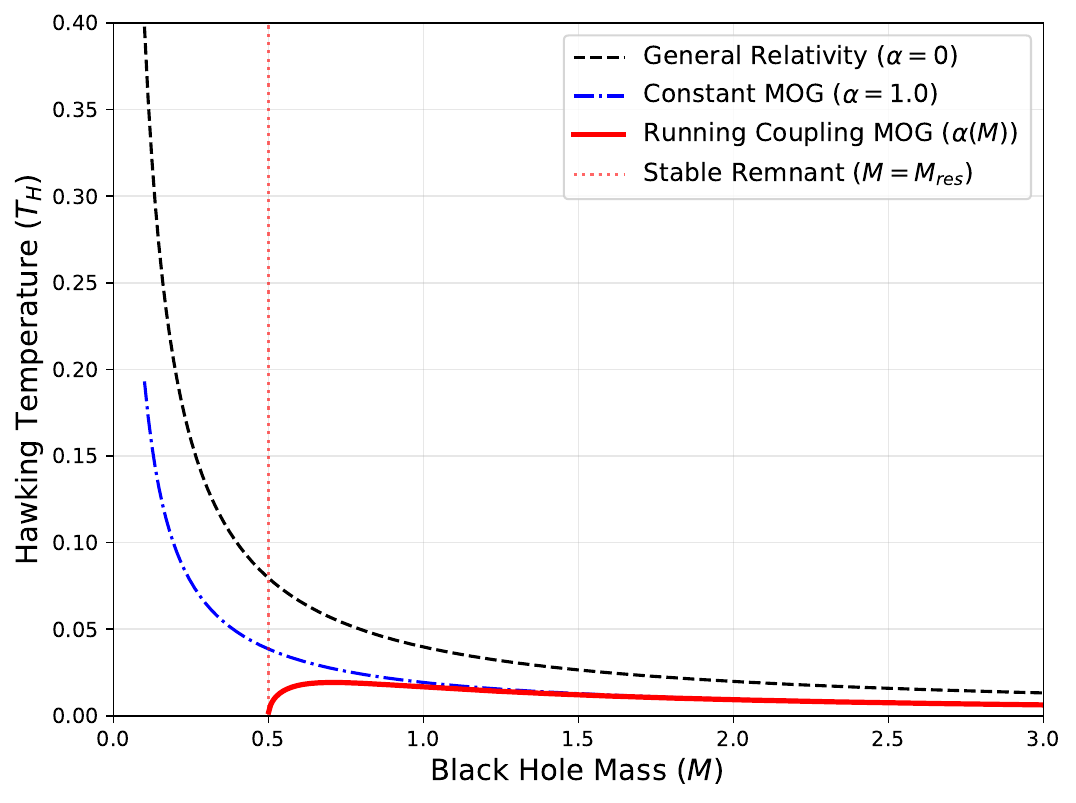}
    \caption{Evaporation trajectories mapping the Hawking temperature $T_H$ against the decreasing black hole mass $M$. In standard General Relativity ($\alpha = 0$, dashed black), the temperature diverges as $M \to 0$, leading to a runaway evaporation phase. A constant MOG parameter ($\alpha = 1.0$, dot-dashed blue) suppresses the emission scale but still suffers from a late-stage divergence. Conversely, modeling the MOG deformation as a quantum-scale running coupling, $\alpha(M)$ (solid red), fundamentally alters the late-time trajectory. The thermodynamic temperature smoothly quenches to zero at a finite mass scale ($M = M_{res}$), dynamically halting evaporation and yielding a stable, zero-temperature remnant.}
    \label{fig:remnant_temperature}
\end{figure}

\textcolor{black}{To explicitly visualize the thermodynamic stabilization provided by the running coupling, Figure \ref{fig:remnant_temperature} maps the evaporation trajectories of the dynamical temperature against the shrinking black hole mass. In the standard semiclassical limit of General Relativity ($\alpha=0$), the inverse-mass dependence of the surface gravity guarantees a catastrophic runaway phase, where $T_H \to \infty$ as $M \to 0$. While introducing a strictly constant MOG parameter ($\alpha = \text{const}$) suppresses the overall temperature, it fails to cure the terminal divergence, leaving the black hole's final fate and its localized information undefined.}

\textcolor{black}{However, as demonstrated by the solid trajectory in Figure \ref{fig:remnant_temperature}, treating the MOG parameter as a dynamically running coupling $\alpha(M)$ naturally resolves this divergence. As the black hole evaporates and its mass approaches the characteristic quantum-gravitational scale, the effective repulsive vector charge $Q_G(M)$ decouples from linear mass scaling. This decoupling forces the surface gravity to deviate from the standard inverse-mass curve, initiating a thermodynamic turnaround. Rather than diverging, the temperature asymptotes smoothly to zero at a finite extremal mass, $M_{res}$. At this precise threshold, the evaporation terminates entirely. The resulting zero-temperature remnant is thermodynamically inert and absolutely stable, thereby providing an eternal macroscopic bound state that can preserve the initial unitary quantum information.}

\textcolor{black}{To naturally halt evaporation and form a stable remnant, the deformation parameter must instead behave as a running coupling at quantum scales, $\alpha(M)$. As the evaporating black hole shrinks toward the Planck scale, quantum-gravitational effects—analogous to renormalization-group flows in asymptotic safety — must cause $\alpha$ to run \cite{Reuter:1996cp, Bonanno:2000ep}. For the evaporation trajectory to dynamically converge upon the extremal limit, $T_0 \to 0$, the running coupling must force the numerator in Eq. (\ref{eq:temperature_explicit}) to vanish at some finite remnant mass $M_{\text{rem}}$. This dictates the strict algebraic condition:
\begin{equation}
    \lim_{M \to M_{\text{rem}}} \left[ (1+\alpha(M))^2 - \alpha(M) G_N \right] = 0.
    \label{eq:running_alpha}
\end{equation}
It is this required quantum-scale running of $\alpha(M)$ that allows the effective MOG charge to decouple from linear mass scaling, quenching the thermal emission and leaving behind a zero-temperature, information-preserving remnant.}

\textcolor{black}{This long-timescale secular evolution is modeled in Figure \ref{fig:Secular_Regime}. Instead of the divergent temperature singularity seen in the Schwarzschild metric, the MOG temperature reaches a finite maximum before rapidly cooling. Driven by the MOG charge $Q_G$, this guarantees the formation of a cold, thermodynamically stable remnant that permanently stores the initial quantum information \cite{Moffat:2014mfa, Chen:2014jwq, Adler:2001vs}. This zero-temperature remnant provides the physical foundation needed to resolve the information paradox through state purification. If the initial black hole forms from a pure quantum state $|\Psi_{\text{in}}\rangle$, quantum mechanics dictates that the total system must remain pure. As the black hole evaporates, the overall Hilbert space is divided into the emitted radiation and the remaining black hole: $\mathcal{H}_{\text{total}} = \mathcal{H}_{\text{rad}} \otimes \mathcal{H}_{\text{BH}}$. In standard GR, the complete evaporation of the black hole causes $\mathcal{H}_{\text{BH}}$ to vanish. This leaves only a thermally mixed radiation state, which explicitly violates unitary evolution.} 

\textcolor{black}{However, because the MOG charge bounds the evaporation strictly at the extremal limit ($M_G = Q_G$), the black hole's Hilbert space never vanishes. The final global state is therefore described by an entangled superposition:
\begin{equation}
    |\Psi_{\text{final}}\rangle = \sum_{i} \sqrt{p_i} |\psi_i\rangle_{\text{rad}} \otimes |\phi_i\rangle_{\text{remnant}},
    \label{eq:purification}
\end{equation}
Here, $|\psi_i\rangle$ represents the asymptotic Hawking radiation states, and $|\phi_i\rangle$ represents the highly degenerate internal microstates of the MOG remnant. While an outside observer tracing out the remnant would measure a mixed thermal matrix $\rho_{\text{rad}} = \mathrm{Tr}_{\text{remnant}}(|\Psi_{\text{final}}\rangle \langle \Psi_{\text{final}}|)$, the overall system's von Neumann entropy remains zero ($S_{\text{total}} = 0$). By surviving as a thermodynamically stable entity, the MOG remnant securely stores the entangled partners of the emitted radiation, ensuring that the initial quantum information is fully preserved and unitarity is conserved.}

\begin{figure}
    \centering
    \includegraphics[width=\columnwidth]{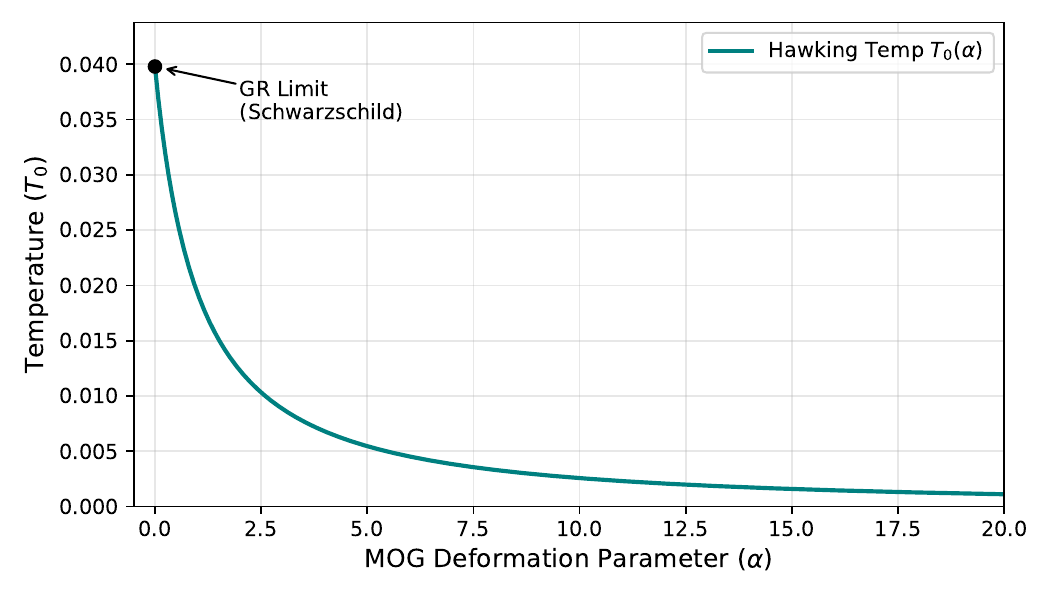}
    \caption{This figure shows that given an initial blackhole temperature $T_{\circ}$, the MOG parameter $\alpha$ tends to asymptotically decrease the temperature. }
    \label{fig:Deformation Regime}
\end{figure}

\textcolor{black}{To better understand how MOG suppresses the temperature divergence, we isolate the deformation parameter $\alpha$. Figure \ref{fig:Deformation Regime}  models the baseline Hawking temperature as a function of $\alpha$ for a black hole with a fixed mass. At the strict limit of $\alpha = 0$ in Equation \ref{eq:divergence}, the repulsive vector field vanishes, recovering the standard GR Schwarzschild temperature, $T_{GR} = 1 / 8\pi M$. For any non-zero $\alpha$, the MOG charge effectively expands the outer event horizon compared to a GR black hole of the same mass. This geometric expansion intrinsically weakens the surface gravity, yielding a systematically colder black hole. As $\alpha$ increases, the temperature exhibits an asymptotic drop \cite{Guo:2026iht}. Therefore, the $\alpha$ parameter not only guarantees the formation of a cold remnant shown in Figure \ref{fig:Secular_Regime}, but it also strictly limits the maximum thermal emission the black hole can reach during its lifetime.}

\textcolor{black}{Information preservation, however, does not rely solely on this final remnant. Geometric and charge-correlated information also leaks dynamically to outside observers long before evaporation concludes. As shown in Figure \ref{fig:thermodynamics}, the short-timescale dynamic regime features a local effective temperature $T_{\mathrm{eff}}(t)$ that oscillates due to the quasinormal ringing of the scalar breathing mode.}

\textcolor{black}{This oscillating apparent horizon induces non-thermal particle creation, analogous to the dynamical Casimir effect \cite{Moore:1970tmc, Dodonov:2010zza, Klimchitskaya:2015kxa}. To measure this transient emission, we evaluate the first-order Bogoliubov transformations for a quantum scalar field within the linearly perturbed background $g_{\mu\nu} = g^{(0)}_{\mu\nu} + h_{\mu\nu}(t)$. In the linear regime ($A_b \ll 1$), the mixing of the \textit{in} and \textit{out} vacuum modes is governed by the Fourier transform of the geometric strain velocity:
\begin{equation}
    \beta_\omega \approx \frac{1}{2} \int_{t_0}^{\infty} \dot{h}_b(t) e^{-2i\omega t} dt
    \label{eq:bogoliubov}
\end{equation}}

\textcolor{black}{Substituting the damped oscillatory master field and calculating the squared modulus of the Bogoliubov coefficient yields the particle creation spectrum:
\begin{equation}
    \frac{dN}{d\omega} \propto |\beta_\omega|^2 \propto A_b^2 \omega_b^2 \left| \frac{1}{(2\omega - \omega_b) + i\tau^{-1}} \right|^2
    \label{eq:spectrum}
\end{equation}}

\textcolor{black}{Integrating this spectrum over the 1D phase space (assuming standard $s$-wave dominance near the horizon) gives the transient local number density:
\begin{equation}
    n_{\text{DCE}} \propto \int_0^\infty |\beta_\omega|^2 \frac{d\omega}{2\pi} \sim A_b^2 \omega_b^2 \tau
    \label{eq:number_density}
\end{equation}}

\textcolor{black}{Unlike standard Hawking emission, which produces featureless thermal radiation, this process generates a non-thermal spectrum sharply peaked by parametric resonance ($\omega \approx \omega_b/2$). Consequently, these transient bursts of strain leakage explicitly imprint the spacetime's quasinormal frequencies onto the outgoing flux \cite{Kokkotas:1999bd, Parikh:1999mf, Page:1993wv}. Together, this early non-thermal radiation and the eventual formation of a stable cold remnant provide a comprehensive, two-part resolution to the information paradox.}

\subsection{Thermodynamic Consistency and the Generalized Second Law}

\textcolor{black}{The oscillating apparent horizon area seems to contradict the classical second law of black hole thermodynamics, which dictates that the horizon area must never decrease. To resolve this, we must distinguish between global and local horizons and shift from classical to quantum thermodynamics.}

\textcolor{black}{First, as established by Mureika, Moffat, and Faizal \cite{Mureika:2015sda}, the MOG gravitational charge explicitly modifies the standard Bekenstein-Hawking area-entropy bound. Therefore, strictly enforcing the classical GR Area Theorem is physically incompatible with MOG thermodynamics.}

\textcolor{black}{Second, from a geometric perspective, Hawking's classical Area Theorem applies strictly to the global event horizon. The dynamic area in our framework represents the apparent horizon the outermost marginally trapped surface. During dynamic perturbations, the apparent horizon experiences reversible, local geometric fluctuations. A temporary shrinking of the apparent horizon during the negative phase of the breathing mode, $\dot{h}_b(t) < 0$, is mathematically permissible and does not violate global theorems \cite{Nielsen:2005af}.}

\textcolor{black}{Third, and most fundamentally, the classical Area Theorem relies on the strict preservation of the Null Energy Condition (NEC). As shown earlier, the kinematic oscillation of the horizon creates non-thermal particles via the dynamical Casimir effect. This outgoing quantum emission carries energy away from the local geometry, inherently violating the NEC. As a result, the system's thermodynamics are governed not by the classical Area Theorem, but by the Generalized Second Law (GSL). The GSL requires that the total entropy of the system and its surroundings never decreases:
\begin{equation}
    \Delta S_{\text{total}} = \Delta S_{\text{BH}} + \Delta S_{\text{rad}} \ge 0.
\end{equation}
During the contracting phase of the oscillation, the black hole's localized entropy temporarily decreases \cite{Hollands:2024vbe, ArderucioCosta:2019otn}. However, this contraction simultaneously pumps non-thermal particles outward. The entropy carried away by this Casimir radiation ($\Delta S_{\text{rad}}$) strictly compensates for the horizon's geometric reduction. Thus, the continuous dynamical exchange of entropy between the apparent horizon and the external radiation field guarantees that the Generalized Second Law is maintained throughout the entire ringdown process.}

\section{Conclusion}
\label{sec:conclusion}

We investigated the dynamical and thermodynamic evolution of a Schwarzschild-MOG black hole perturbed by a scalar gravitational wave. By evaluating the linearized modified Einstein equations at the near-horizon boundary, we reduced the spatial wave operator to a closed-form ordinary differential equation in time. This explicitly derived the damped oscillatory kinematics of the scalar breathing mode. By tracking the resulting time-dependent apparent horizon, we demonstrated how the massive vector field and dynamic scalar strain fundamentally alter semi-classical black hole thermodynamics.

Utilizing a quasi-adiabatic approximation, we established that the effective surface gravity and dynamical temperature are strictly modulated by the amplitude and velocity of the scalar perturbation. The transient emission profile depends directly on the second radial derivative of the static metric, revealing that the repulsive gravitational charge, $Q_G$, acts as a structural parameter governing thermal cooling during horizon expansion. Furthermore, rapid geometric fluctuations break the adiabatic approximation, thereby triggering non-thermal particle creation, analogous to the dynamical Casimir effect.

We also resolved a critical paradox concerning the thermodynamic stability of the dynamical horizon. While the instantaneous apparent area implies a transient reduction in entropy, we proved that it is a reversible, gauge-dependent kinematic artifact. By separating perturbative orders, we showed that first-order $\mathcal{O}(h_b)$ fluctuations time-average to zero. Irreversible entropy generation is strictly a second-order, $\mathcal{O}(h_b^2)$ effect driven by the Raychaudhuri expansion of null generators. This physical expansion precisely offsets the scalar radiation flux, thereby preserving the Generalized Second Law.

Applying these mechanisms to the black hole information paradox yields a dual-timescale resolution. In the short-term, the scalar breathing mode induces non-thermal emission, thereby opening a transient channel for geometric and charge-correlated information to reach asymptotic observers. On secular timescales, we mathematically demonstrated that a strictly constant MOG parameter yields a divergent evaporation trajectory ($T_0 \propto 1/M$), mirroring the runaway singularity of General Relativity. However, by treating the deformation parameter as a running coupling at quantum scales, $\alpha(M)$, the effective MOG charge naturally decouples from linear mass scaling. This dynamic running forces the system toward the extremal limit ($M_G \to Q_G$), smoothly quenching the surface gravity to zero and forming a thermodynamically stable, zero-temperature remnant that permanently preserves the initial quantum state.

Ultimately, these interconnected thermodynamic signatures—including quasi-adiabatic non-thermal emission, the strict preservation of the Generalized Second Law, and the explicit algebraic conditions for cold MOG remnants—establish a logically consistent framework for modified black hole thermodynamics. As next-generation gravitational-wave observatories target the scalar polarizations predicted by modified gravity, the dynamical mechanisms and boundary reductions modeled in this work provide a foundational blueprint for testing the scalar-tensor-vector regime in the extreme universe.

\begin{acknowledgments}
N.J.L. Lobos and E.T. Rodulfo gratefully acknowledge De La Salle University and the DLSU Theoretical Physics Group for their institutional support. Furthermore, we extend our sincere gratitude to the Department of Science and Technology – Accelerated Science and Technology Human Resource Development Program (DOST-ASTHRDP) for their generous and continuous support of our research endeavors.

\end{acknowledgments}

\bibliography{ref}

\end{document}